\documentclass[
aps,%
12pt,%
final,%
notitlepage,%
oneside,%
onecolumn,%
nobibnotes,%
nofootinbib,%
superscriptaddress,%
]%
{revtex4}
\usepackage{epsfig}
\usepackage{graphics}

\newcommand{\be}{\begin{equation}}
\newcommand{\ee}{\end{equation}}
\newcommand{\bn}{\begin{eqnarray}}
\newcommand{\en}{\end{eqnarray}}
\usepackage[english]{babel}
\usepackage{epstopdf}
\usepackage{dcolumn}
\usepackage{bm}
\usepackage{units,nicefrac}
\usepackage{natbib}

\begin{document}
	
	\title{Change in the Crystallization Features of Supercooled Liquid Metal with an Increase in the Supercooling Level}
	
	\author{\firstname{B.\,N.}~\surname{Galimzyanov}}\email{bulatgnmail@gmail.com}
	\affiliation{Kazan Federal University, 420008 Kazan, Russia}
	
	\author{\firstname{D.\,T.}~\surname{Yarullin}}
\affiliation{Kazan Federal University, 420008 Kazan, Russia}
	
	\author{\firstname{A.\,V.}~\surname{Mokshin}}
	\affiliation{Kazan Federal University, 420008 Kazan, Russia}

	\begin{abstract}

The process of homogeneous crystal nucleation has been considered in a model liquid, where the interparticle interaction is described by a short-range spherical oscillatory potential. Mechanisms of initiating structural ordering in the liquid at various supercooling levels, including those corresponding to an amorphous state, have been determined. The sizes and shapes of formed crystal grains have been estimated statistically. The results indicate that the mechanisms of nucleation occurs throughout the entire considered temperature range. The crystallization of the system at low supercooling levels occurs through a mononuclear scenario. A high concentration of crystal nuclei formed at high supercooling levels (i.e., at temperatures comparable to and below the glass transition temperature $T_g$) creates the semblance of the presence of branched structures, which is sometimes erroneously interpreted as a signature of phase separation. The temperature dependence of the maximum concentration of crystal grains demonstrates two regimes the transition between which occurs at a temperature comparable to the glass transition temperature $T_g$.
\end{abstract}

\maketitle

In terms of thermodynamics, a supercooled liquid is in a state of unstable equilibrium, which results in the appearance of domains of a crystal phase in it~\cite{Kashchiev_Nucleation_2000,Kelton_2010,Jetp_2012,Tovbin_2014,Malenkov_2012,Egami_Levashov_2010}. At the same time, the character of the process of structural ordering should significantly depend on the conditions under which the supercooled state was formed, in particular, on the cooling rate of the liquid and on its final supercooling level $\Delta T/T_m=1-T/T_m$, where $T_m$ is the melting temperature of the system~\cite{Kelton_2010,Fomin_2012,Fokin_Zanotto_2006}. At low and moderate supercooling levels covering the temperature range $T_g<T<T_m$, crystallization is usually initiated through the scenario of crystal nucleation, which is described within classical nucleation theory~\cite{Kashchiev_Nucleation_2000,Kelton_2010,Fokin_Zanotto_2006}. At temperatures below the glass transition temperature $T_g$, which correspond to high supercooling levels, the system forms an amorphous (glassy) state, where the space-time scales of crystal nucleation are beyond the sensitivity range of modern experimental instruments. Furthermore, an increase in the supercooling level of the amorphous system leads to an increase in the concentration of small crystal grains, the characteristics of the formation and growth of which cannot be predicted/described within classical nucleation theory~\cite{Kashchiev_Nucleation_2000}. For this reason, a commonly accepted complete understanding of the crystallization process of amorphous systems is still absent in spite of numerous studies~\cite{Sosso_Michaelides_2016,Peng_Wang_2015,Ivanov_Fedorov_2014,Tovbin_2014,Mokshin_Galimzyanov_PCCP_2017}. In particular statements of the possibility of crystallization of liquids at high supercooling levels through the mechanism of phase separation are contradictory~\cite{Malek_Morrow_2015,Trudu_Parrinello_2006,Bartell_Wu_2007}. Some authors present signatures that can be considered as indications of the spinodal mechanism of structural ordering in single-component supercooled liquids~\cite{Trudu_Parrinello_2006}, whereas other authors~\cite{Bartell_Wu_2007,Skripov_1974} argue that the spinodal mechanism is impossible in these systems. The aim of this work is to consider this issue.

\begin{figure}[ht]
	\centering
	\includegraphics[width=0.6\linewidth]{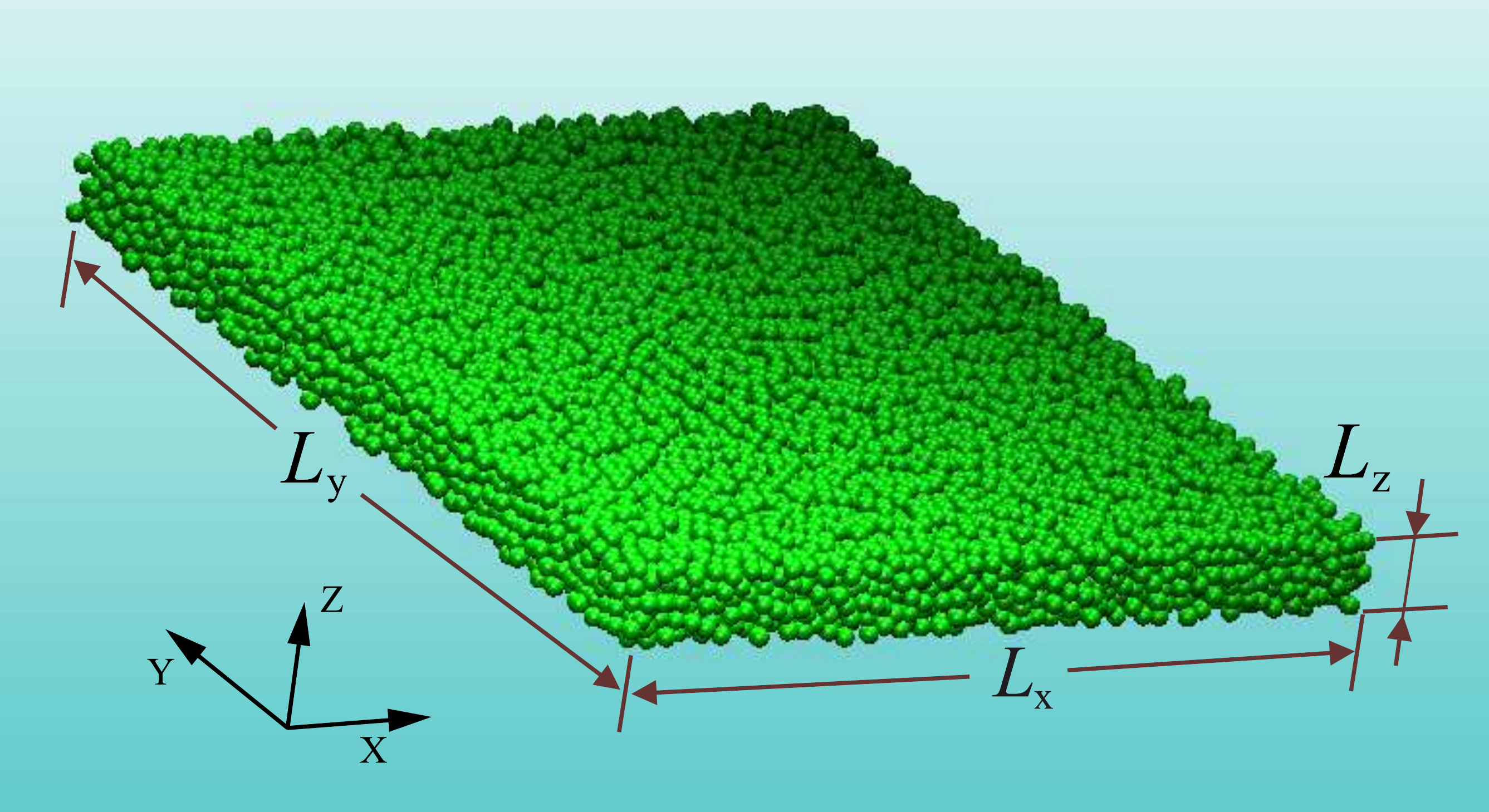}
	\caption{(Color online) Simulated system with the volume $V=L_{x}\times L_{y}\times L_{z}$, where $L_{x}=L_{y}\approx11.5L_{z}$ and $L_{z}\simeq4.8\,\sigma$. The number density of particles of the system is $\rho=N/V=0.936\,\sigma^{-3}$.}\label{fig_1}
\end{figure}

We consider a multiparticle system, where the interaction between particles is described by a shorty-range spherical oscillatory-potential~\cite{Dzugutov_1992,Roth_Denton_2000}, which effectively reproduces the ion-ion interaction in metal melts. This specific potential promotes the formation of a relatively stable amorphous state. The simulated system is shown in Fig.\ref{fig_1}. We consider the temperature range $ T = (0.5-1.4)\,\epsilon/k_B$ on the isobar $p=15\,\epsilon/\sigma^3$, which corresponds to temperatures below the melting temperature of the system $T_m\simeq1.72\,\epsilon/k_B$ and supercooling levels from $\Delta T/T_m\approx0.19$ (at $T=1.4\,\epsilon/k_B$) to $\approx0.71$ (at $T=0.5\,\epsilon/k_B$)~\cite{Roth_Denton_2000}. \footnote{Physical quantities are given in Lennard-Jones units: the effective diameter of the particle $\sigma$, the energy unit $\epsilon$, the time unit $\tau=\sigma\sqrt{m/\epsilon}$, where $m$ is the mass of the particle; the temperature $T$ and pressure $p$ are measured in units of $\epsilon/k_{B}$ and $\epsilon/\sigma^{3}$, respectively, where $k_{B}$ is the Boltzmann constant.}  The glass transition temperature of the system is $T_g\simeq0.78\,\epsilon/k_B$ (at the cooling rate \,$0.04\epsilon/(k_BT)$ on the isobar $p=15\epsilon/\sigma^3$). Such high supercooling levels of the single component melt are now available by existing experimental methods. In particular, experimental results for amorphous titanium cooled at a rate if $~10^{14}$ K/s to the supercooling level $\Delta T/T_m\approx0.85$ were reported in~\cite{Zhong_Wang_2014}.

The structural analysis and identification of crystal grains were performed through the calculation of the local orientational order parameters, which were introduced in~\cite{Steinhardt_1983,Mickel_Mecke_2013}, and by means of the algorithm proposed in~\cite{Wolde_Frenkel_1996}. According to~\cite{Wolde_Frenkel_1996}, each particle incoming to a crystal nucleus can contain no less than seven neighbors. The critical size $n_c$ of the nucleus is estimated through the statistical treatment of the growth trajectories of the first (largest) nucleus that are obtained for independent simulated samples. Details of the algorithm can be found in~\cite{Mokshin_Galimzyanov_JPCS_2012,Mokshin_Galimzyanov_JCP_2014,Mokshin_Galimzyanov_2015}.

\begin{figure*}[ht]
	\centering
	\includegraphics[width=1.0\linewidth]{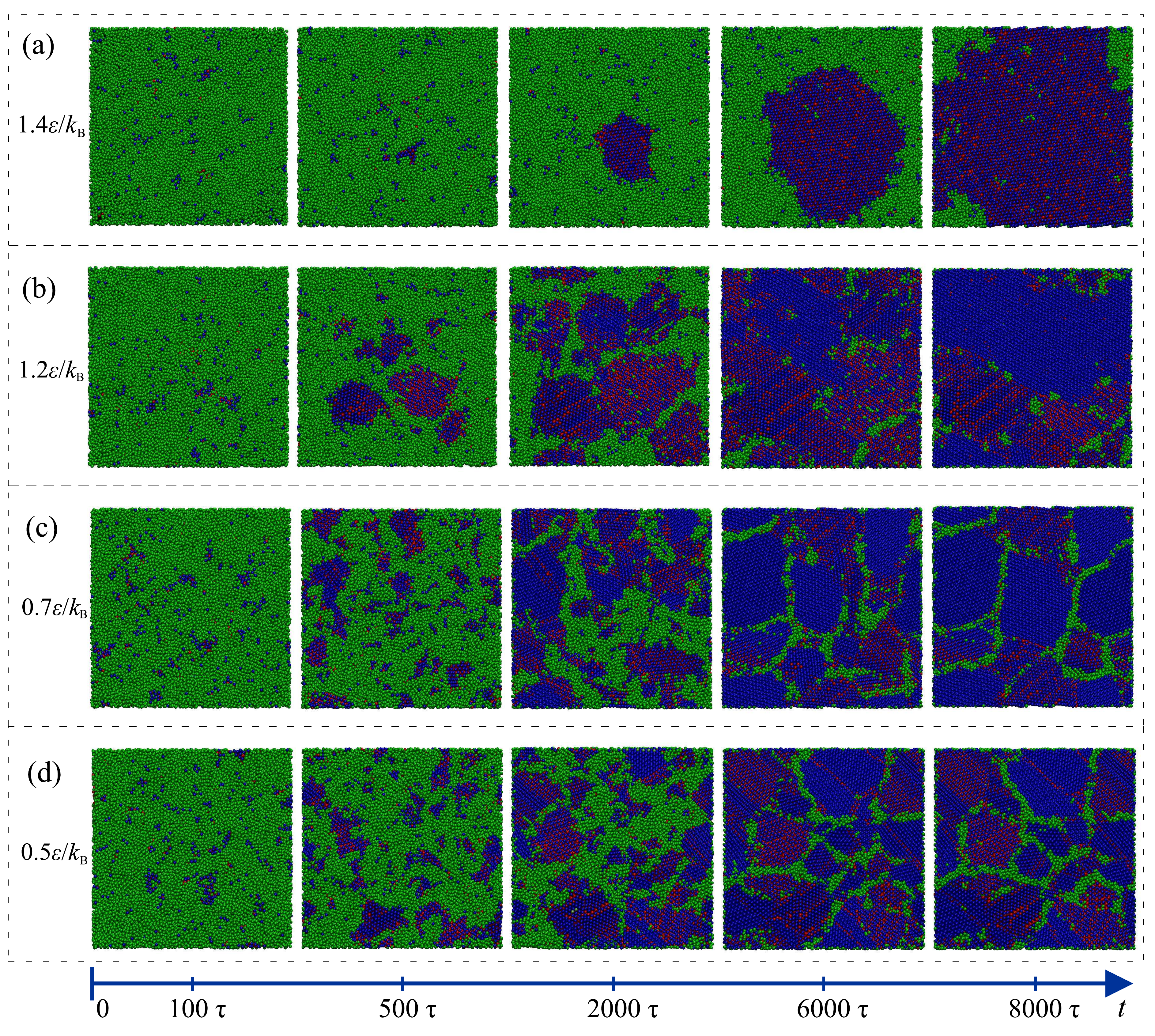}
	\caption{(Color online) Configurations of the system at different times and temperatures $T=$(a) $1.4\,\epsilon/k_{B}$, (b) $1.2\,\epsilon/k_{B}$, (c) $0.7\,\epsilon/k_{B}$ and (d) $0.5\,\epsilon/k_{B}$. Dark blue points mark particles of the fcc structure for which the parameters of the local orientational order are $q_{4}\simeq0.19$, $q_{6}\simeq0.578$ и $q_{8}\simeq0.404$~\cite{Malek_Morrow_2015}. Particles of the hcp structure are identified at the parameters $q_{4}\simeq0.097$, $q_{6}\simeq0.485$ and $q_{8}\simeq0.317$ and are shown in dark red. Particles of the disordered phase are given in light green.}\label{fig_2}
\end{figure*}

Regions containing ordered structures in systems at various supercooling levels at different times from the time of preparation were identified by the structural analysis. In particular, Fig.\ref{fig_2} shows configurations of a system obtained for different times at temperatures $T=0.5\,\epsilon/k_B,  0.7\,\epsilon/k_B, 1.2\,\epsilon/k_B$, and $1.4\,\epsilon/k_B$. The temperatures $T=0.5\,\epsilon/k_B$ and $0.7\,\epsilon/k_B$ are below the glass transition temperature $T_g\simeq0.78\,\epsilon/k_B$ and correspond to an amorphous system. As is seen in Fig.\ref{fig_2}, at all temperatures in the initial stage of crystallization, i.e., in the time interval $t\in[0,300]\,\tau$, the process of spontaneous formation of small crystal nuclei containing less than $50$ particles is observed. These nuclei are unstable and are solved in the volume of the initial "parent" phase. Nuclei with the critical size $n_c$ capable of growth are formed at times $t>300\,\tau$. The critical size $n_c$ is calculated from the analysis of distributions of average times of appearance of a crystal nucleus with a certain size[$12,26$]. It was found that, with an increase in the supercooling level, the critical size decreases from $n_c\simeq83$ (at $T=1.4\,\epsilon/k_B$) to $n_c\simeq67$ (at $T=0.5\,\epsilon/k_B$) particles, which is consistent with classical theory [$1,2$]. It is remarkable that the variation of the critical size in such a wide temperature range is more than ten particles, but is insignificant.

According to Fig.\ref{fig_2}a, the crystallization of the system at low supercooling levels (e.g., at the temperature $T=1.4\,\epsilon/k_B$) occurs through the formation and growth of a single nucleus whose shape is quite smooth. Such a feature of nucleation is well known in classical theory and is typical of the mononuclear nucleation scenario~\cite{Kashchiev_Nucleation_2000,Kelton_2010}. With an increase in the supercooling level of the system to $\Delta T/T_m\simeq0.3$, which corresponds to the temperature $T=1.2\,\epsilon/k_B$, crystallization begins to occur through the so-called polynuclear mechanism~\cite{Kashchiev_Nucleation_2000}, at which the concentration of nuclei with a supercritical size increases quite rapidly. It is seen in Fig.\ref{fig_2}b that such mechanism is accompanied by the coalescence of nuclei. This leads to the complete crystallization of the system through the formation of a single crystal with a small number of dislocations. It is remarkable that the coalescence of nuclei occurs according to the \emph{oriented attachment model} described in detail in~\cite{Ivanov_Fedorov_2014,Fedorov_Osiko_2014}; i.e., coalescence occurs through the displacements and rotations of nuclei with respect to each other~\cite{Ivanov_Fedorov_2014}.

\begin{figure}[ht]
	\centering
	\includegraphics[width=0.85\linewidth]{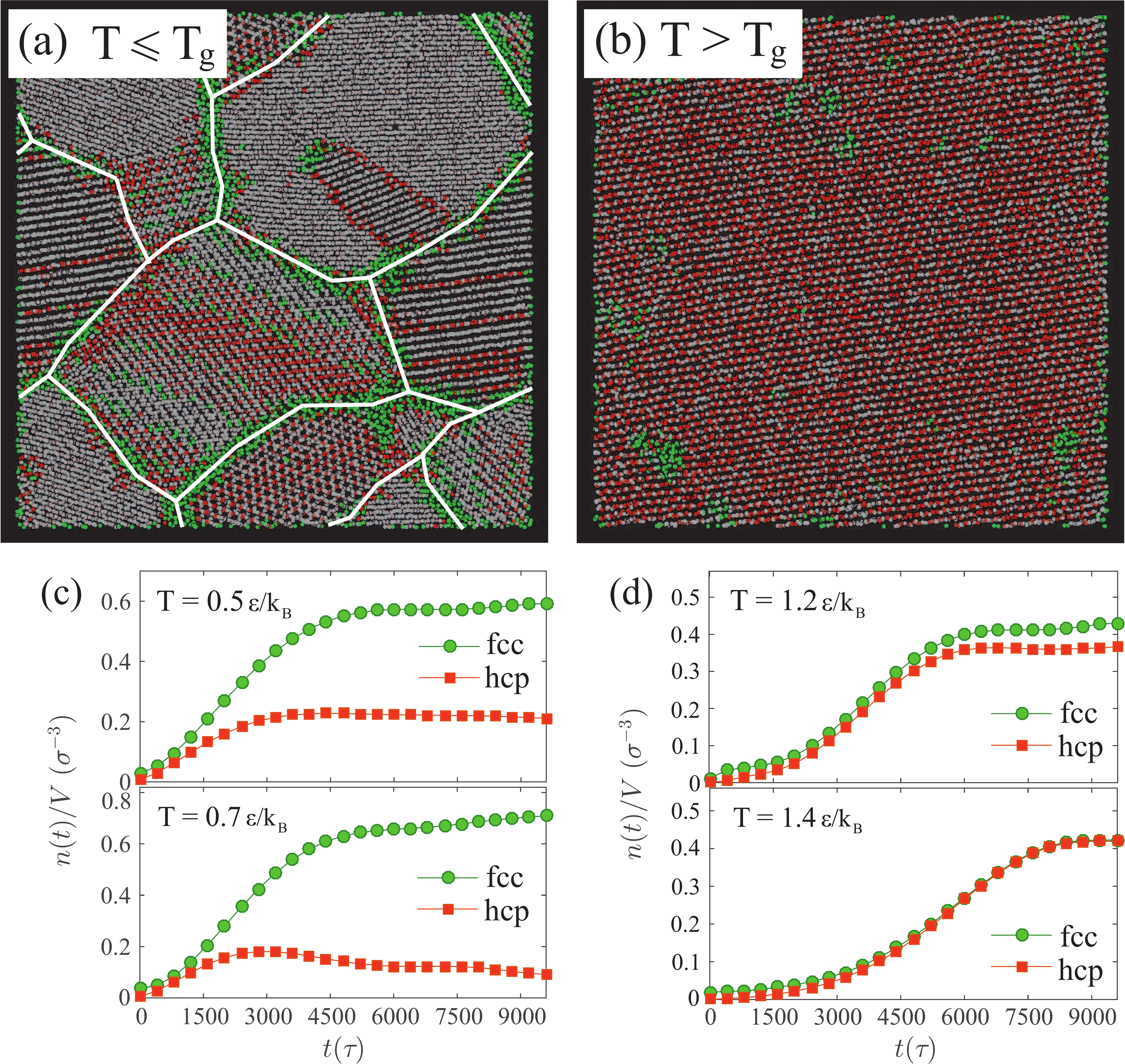}
	\caption{(Color online) (Upper panels) Configurations of the system at the final stage of crystallization at temperatures (a) $T<T_{g}$ and (b) $T>T_{g}$. Particles forming the disordered phase are shown by green circles. Particles of the crystal phase with the fcc and hcp lattices are given by white and red circles, respectively. White lines are boundaries between different crystal domains. (Lower panels) Time dependences of the density of particles of the fcc and hcp phases at temperatures (upper c) $0.7\,\epsilon/k_{B}$, (lower c) $T=0.5\,\epsilon/k_{B}$, (upper d) $T=1.4\,\epsilon/k_{B}$, and (lower d) $1.2\,\epsilon/k_{B}$.}
	\label{fig_3}
\end{figure}

At high supercooling levels, structural ordering proceeds as follows. In particular, a high concentration of small crystal grains is observed in the initial stage of crystallization at temperatures $T\in[0.5\epsilon/k_B, 0.8\,\epsilon/k_B$]. This creates the semblance of the presence of branched structures, which was erroneously interpreted in~\cite{Trudu_Parrinello_2006,Bartell_Wu_2007} as a signature of phase separation. As is seen in Figs.\ref{fig_2}c and \ref{fig_2}d, the incomplete coalescence of these nuclei results in the formation of fragmented structures containing dislocations, which prevent the formation of a single crystal. As a result, a polycrystalline structure is formed. As an example, Fig.\ref{fig_3}a, the system is as ensemble of crystalline domains with different orientations of the plane of the crystal lattice. On the other hand, polycrystalline structures are not formed at low and moderate supercooling levels (i.e., at $T>T_g$). In particular, it is seen in Fig.\ref{fig_3}b that the system is an almost perfect single crystal.

It is noteworthy that the observed that the observed crystallization of the amorphous system at high supercooling levels $\Delta T/T_m$ reaching $\approx0.71$ is surprising at first glance. Indeed, the velocities of particles are very low at the temperature $T=0.5\,\epsilon/k_B$ corresponding to this supercooling. Nevertheless, estimates show that the critical size of nucleus is $~67$ particles and the rate of transition of particles to the crystal phase is $g^+_{n_c}\simeq(18\pm4)\,\tau^{-1}$. The recalculation of this $g^+_{n_c}$ value to the case of iron with the parameters of potential $\sigma\approx2.517$\,\AA and $\epsilon\simeq16.15$\,kcal/mol~\cite{Bykata_Borges_2005} gives $g^+_{n_c}\approx10^{14}s^{-1}$. Consequently, the crystal nucleus reaches a linear size of $~1 cm$ in a time interval of $50-80$ yr. Such a time scale attributed to the crystallization of metal systems seems quite correct~\cite{Zhong_Wang_2014}.

In order to determine the type of symmetry of the crystal lattice of the formed ordered phase, we calculated the parameters of the local orientational order~\cite{Mickel_Mecke_2013}. The found parameters indicate that crystal phases with face-centered cubic (fcc) and hexagonal close-packed (hcp) lattices prevail in the system (see Fig.\ref{fig_2}). The structural analysis allowed the calculation of the time dependences characterizing the number densities of particles $n_{fcc}(t)/V$ and $n_{hcp}(t)/V$ in the fcc and hcp structures, respectively (see Figs.\ref{fig_3}c and \ref{fig_3}d). The results show that the relation between these densities significantly depends on the supercooling level.
In particular, at temperatures $T<T_g$, the fraction of particles forming the hcp structures is much smaller that the fraction of particles in the fcc structures. The difference between these fractions increases with the supercooling level.

\begin{figure}[ht]
	\centering
	\includegraphics[width=0.85\linewidth]{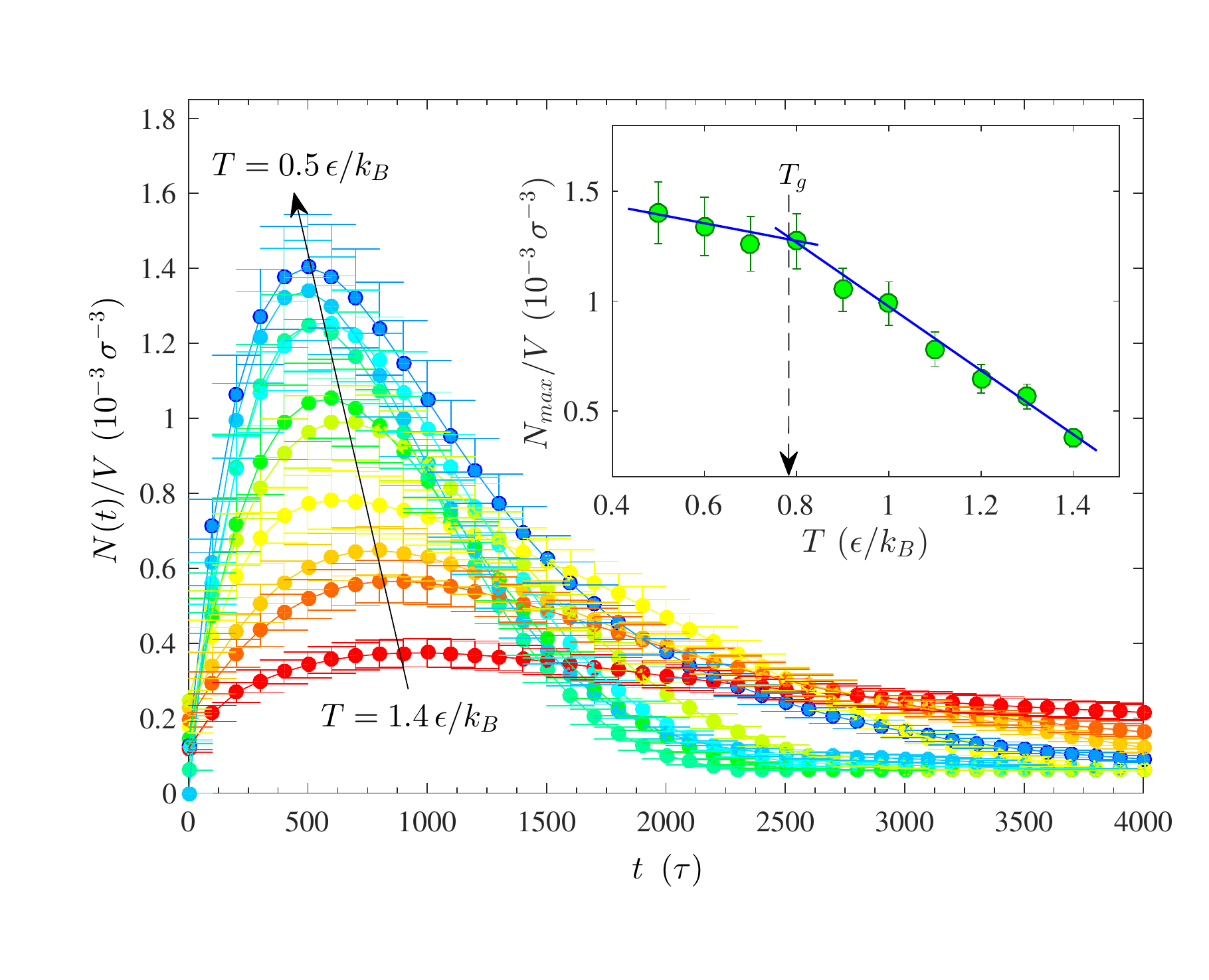}
	\caption{(Color online) Time dependences of the concentration of crystal nuclei with a supercritical size, $N(t)/V$, calculated for a various temperatures of the system. The inset shows the temperature dependence of the maximum concentration of crystal grains $N_{max}/V$. Two regimes a transition between which occurs at a temperature comparable to the glass transition temperature $T_g$ are seen.}
	\label{fig_4}
\end{figure}

Figure \ref{fig_4} shows time dependences of the concentration of crystal nuclei $N(t)/V$ , calculated for various temperatures. These dependences $N(t)/V$ are similar and exhibit a pronounced maximum. The position of the maximum on the time scale separates the nucleation regime (small times, where $N(t)/V$ increases) and a regime associated with the process of growth and coalescence of crystal grains (times at which $N(t)/V$) decreases).
 As is seen in Fig.\ref{fig_4}, the height of the maximum in $N(t)/V$, which characterizes the maximum concentration of grains $N_{max}/V$, strongly depends on the temperature. The inset of Fig.\ref{fig_4} shows the temperature dependence of the maximum concentration of grains $N_{max}/V$. This temperature dependence exhibits two pronounced regimes. In particular, $N_{max}/V$ decreases linearly with the temperature in the temperature interval $T\in[0.8\epsilon/k_B,\,1.4\,\epsilon/k_B$]. However, at temperature below $T=0.8\,\epsilon/k_B$, the concentration of nuclei $N_{max}/V$ weakly depends on the temperature of the system: $N_{max}/V$ increases insignificantly with reproduction of the temperature. It is remarkable that the point of intersection of linear sections in the temperature dependence $N_{max}(T)/V$ almost coincides with the glass transition temperature of the system $T_g\,\simeq0.78\epsilon/k_B$. This surprising nontrivial result requires a more detailed analysis and test in application systems with another characteristic interparticle interaction (polymers, colloidal systems, etc).

\vspace{0.5cm}

We are grateful to Prof. S.V. Demishev and V.V. Glushkov (Prokhorov Genereal Physics Institute, Russian Academy of Sciences, Moscow) for stimulating discussions and recommendations. This work was supported by Kazan Federal University and the Russian Foundation for Basic Research (project No. 18-32-00021) and by the Ministry of Education and Science of the Russian Federation (subsidy for state assignment in science no. 3.2166.2017/4.6 to Kazan Federal University). The molecular dynamics calculations were performed with the use of equipment of the Supercomputer Complex, Moscow State University, and at Computer Clusters, Kazan Federal University.

\bibliographystyle{unsrt}

\end{document}